\documentclass[preprint,showpacs,amsmath,amssymb]{revtex4}

\usepackage{graphicx}
\usepackage{dcolumn}
\usepackage{bm}

\begin{document}

\title{What evidences the mass distribution in halo nucleus?}

\author{A.\ I.\ Steshenko}
\affiliation{Bogolyubov Institute for Theoretical Physics of the NAS of Ukraine, \\
Metrolohichna Str., 14b, Kyiv-143, 03143, Ukraine}

\date{\today}

\begin{abstract}
Detailed analysis of the distribution of nuclear matter in the
halo nucleus $^{11}$Be is performed within the microscopic approach based
on the flexible variational basis of the functions of the "polarized"\,
orbitals model. It is shown that the proposed approach reproduces, in
agreement with the experiment, the most distinguishing features of the
mass distribution in this system. Taking into account the results of
specific calculations and the exceptionally high reliability of the
experimental data on the nuclear density distribution in the "tail"\, of
$^{11}$Be, we make a conclusion about the existence of new structures
with the sizes considerably exceeding the size of the nucleus (up to tens
of $fm$) and the density being thousands times less that the typical
nuclear one.
\end{abstract}


\maketitle

\section{Introduction}

As is known, during the last 10-15 years, a number of nuclides with the
anomalous N/Z ratio has been discovered (see, for instance, ~\cite{1},
~\cite{2}, ~\cite{3}, ~\cite{4}, ~\cite{5}, ~\cite{6}). These nuclei are
far from the beta-stability area and may closely approach the boundary of
the nuclear or nucleon stability. In most of the cases, these are the
strongly neutron-excessive isotopes of light and middle elements, which
include a number of exotic nuclei with interesting and unusual properties
and structure. Naturally, these nuclides attracted considerable attention
of both experimentalists and theorists.

Let us say a few words about the term "halo nuclei". The latter has
become very popular in recent years and is also used for the well known
nuclides (like $^{6}$He, $^{8}$He, etc.). For this reason, in order to
bring, at least partially, the terminology and the physical meaning into
agreement, there has been one more extravagant term, "Borromean nuclei",
introduced ~\cite{7}. Borromean or, alternatively, halo nuclei of II-nd
kind, represent weakly coupled and very much fragile nucleon systems
being at the same time rather compact in size. They should be more
exactly referred to as "neutron skin"\, rather than "halo"\, nuclei.

Quite different situation is observed in the case of halo nuclei of I-st
kind (the one-neutron halo nuclei $^{11}$Be and $^{19}$C). These nuclides
are unusual from every point of view.  Let us mention the anomalously
high cross-sections of the interaction with various nuclei, the
unexpected inverted spectra of the lower energy levels, and the most
unexpected mass distribution with the compact core nucleus in the center
accompanied by a long sharply directed "tail"\, of very thinned nuclear
matter  ~\cite{1}, ~\cite{8}.

We are not going to consider in detail the basic features of halo nuclei,
since there is a number of exhaustive reviews on the subject in the
literature (see, for instance,  ~\cite{2}, ~\cite{3}, ~\cite{4},
~\cite{5}, ~\cite{6}). Let us cite only the experimental results on the
distribution of the nuclear matter in $^{11}$Be keeping in mind that
these latter will be the major subject of our further theoretical
consideration. These results have been obtained by a large team of
Japanese physicists in RIKEN ~\cite{2}, where the dependence of the
density distribution for the nuclear matter in the nucleus $^{11}$Be on
the distance from the nucleus center has been studied. The experimental
points obtained there have been put within some stripe area bounded by a
bold line, in order to make allowance for the possible inaccuracies in
the determination of density (see Fig.2 in ~\cite{8}). The authors' very
convincing and rather strong statements are cited below.

In first place, the existence of the halo tail in $^{11}$Be nucleus is
associated with the last  weakly coupled neutron only, and, in addition,
this valence neutron is the main material provider for the "tail"\,
density distribution of nuclear matter. The contributions of the rest of
nucleons to the "tail"\, are negligible. The magnitude of the density in
the halo tail is more than three orders less than the typical density of
nuclear matter.

In second place, while in determination of the central and intermediate
(with respect to $r$) regions there is a possibility of effects of
additional factors, which, in principle, may bring the magnitude of
density beyond  the area of experimental data obtained by the authors,
the slope and the magnitude in the "tail"\, distribution are determined
very reliably and accurately. The question is, what are the reasons for
such confidence of the authors? The point is that the most sensitive
process with respect to the distribution of the matter in the "tail" is
that of $^{11}$Be and $^{10}$Be fragmentation, and the cross section
$\sigma_{F}$ of this process is anomalously high. As a result, the
experimental determination of the above mass distribution in $^{11}$Be
can be made with such an impressive accuracy.

One more point important for the theory, which has been emphasized in the
paper ~\cite{8}, can be concisely formulated as the indifference of the
halo tail in relation to the core nucleus. In other words, the presence
of the halo tail in $^{11}$Be almost does not affect the magnitude of the
basic parameters of the core nucleus $^{10}$Be.

All the above mentioned relates to the experimental situation on halo
nuclei. As regards the theory, the mainstream of  creative ideas here is
focused on the cluster model. There are hundreds of papers in the
literature, where the successful description of exotic halo nuclei is
based on the employing one or another of cluster modes, e.g., the core
plus one valence neutron, or the core plus two valence neutrons. At first
sight, such interpretation appears very attractive.  However, a closer
examination reveals here serious difficulties. Let us point out some of
them. To begin with recall the size parameters of $^{11}$Be. The
root-mean-square radius of the nuclear matter in $^{11}$Be is
$R^{m}_{rms}=(2,90\pm0,05)\,fm$ ~\cite{9},~\cite{10}, whereas the radius
of the core nucleus $^{10}$Be is $R^{m}_{rms}=(2,30\pm0,02)\,fm$. In
order to explain the experimental spectrum of the Coulomb decay, the
value of the root-mean-square radius of the valence nucleon and the
diameter of the halo tail must be, in accordance with ~\cite{11},
$R^{m}_{rms}\sim 7\, fm$, and
 $ D_{halo}\approx30\,fm$, respectively. If one assumes
that the valence nucleon moves outside the core (the concept underlying
the cluster representation), a serious question arises as to the origin
of the long-range interaction between the valence neutron and the
nucleons of the core. To treat this problem, one may introduce the
so-called "shallow neutron-core potential"\,~\cite{12}, and, by
appropriately adjusting the parameters for this potential, achieve
success in calculating specific physical quantities. It is clear that
such an interpretation is rather phenomenological, and the situation here
could be hardly said to be satisfactory.

The situation with halo nuclei could be clarified within the consistent
microscopic approach ~\cite{13} based on the many-particle
Schr$\ddot{o}$dinger equation for the wave function  $ \Psi(1,2,...,A)$
with   some particular choice of the "fundamental"\, pair NN-potential.
Actually, the dependence of the wave function  $ \Psi(1,2,...,A)$ on the
coordinates of all A nucleons enables one to calculate almost arbitrary
nuclear property and to reveal the underlying physical picture by means
of the analysis of relevant theoretical and experimental quantities.

Taking into account all the above mentioned, we perform in this work
exactly this analysis for the case of one-neutron halo nucleus $^{11}$Be.
In Section II of the paper, we give a brief presentation of the algorithm
for calculations within the framework of the polarized orbitals model
~\cite{14} taking an accurate account of the hierarchy of basic
principles of the theory. In Section III, we present the results of
calculations of the nuclear matter density in the nucleus $^{11}$Be. The
discussion of theoretical data, along with comparison between the known
experimental data and similar results of other authors, is given in
Section IV.

\section{Many-particle wave function and the variational principle}

\qquad In a microscopic model of interacting particles, the starting
point is the relevant many-particle Hamiltonian, where, as a rule, only
pair interactions are taken into account, i.e.

{\normalsize
\begin{equation}
\label{1}  \mathbf{H}_A\Psi= E\Psi; \qquad \mathbf{ H}_A =
-\sum_{j=1}^{A}\frac{\hbar^2}{2m}\mathbf{\nabla}^2_j - \mathbf{T}_{c.m.}
+ \sum_{i>j=1}^{A}V_{i,j} + U_{Coulomb}\,.
\end{equation} }
 As regards the work ~\cite{14}, which provides the basis for our further
consideration, the pair inter-nucleon interaction employed therein is
taken in the form of effective central exchange NN-potential with the
radial dependence specified by the superposition of  five Gaussian terms
~\cite{15}
{\normalsize
\begin{equation}
 \label{2}
 V_{ij}= \sum_{S,T=0,1} \sum_{\nu =1}^{5}V_{2S+1,2T+1}^{[\nu]} \, \exp
 \{-\frac{(\mathbf{r}_i-\mathbf{r}_j)^2}{\mu^2_{\nu}}\}\cdot \,
 \mathbf{P}_{S,T}(i,j),
\end{equation} }
where  $V_{2S+1,2T+1}^{[\nu]}$ and  $\mu_\nu$ are the intensity and the
radius of the  $\nu$ -th component of the interaction potential between
the $i$-th and $j$-th nucleons, respectively;
 $\mathbf{P}_{S,T}$ is the known projection operator (see, for
instance, ~\cite{16}), which separates out the two-nucleon states with
the spin $S$ and isospin $T$ in the wave function $\Psi$.

Along with the nuclear interaction, the Hamiltonian (1) includes also the
Coulomb repulsion between the protons in the nucleus, which can be
conveniently written in the form of the integral representation
{\normalsize
\begin{equation}
 \label{3}
 U_{Coulomb}=\sum_{i>j=1}^Z \frac{e^2}{|\mathbf{r}_i - \mathbf{r}_j |}
  =\frac{2e^2}{\sqrt{\pi}}\sum_{i>j=1}^Z \int\limits_{0}^{1}
  e^{-\frac{(\mathbf{r}_i-\mathbf{r}_j)^2}{\mu_c^2}} \frac{d\tau}{(1-\tau^2)^{3/2}},
  \quad \mu_c\equiv \frac{1}{\tau}\sqrt{1-\tau^2}.
\end{equation} }
Thus, formally, we deal with the conventional many-particle problem for
bound states. In view of  the possibilities of modern computers, it is
worth to seek the solution to the problem (1)-(3) on the basis of the
well known variational Ritz' principle (see, for instance, ~\cite{17}).
In doing so, it is natural to describe the nuclear system in terms of
individual particle states with the total wave function of the nucleus
being approximated by the antisymmetrized product of one-particle
functions
{\normalsize
\begin{equation}
 \label{4}
\Psi_{E}(1,2,...,A)=\frac{1}{\sqrt{A!}}\det[\varphi_i(j)]; \quad
\varphi_j(i)=\psi_{\mathbf{n}_i}(\mathbf{r}_j;a_i,b_i,c_i)\cdot
\chi_{\frac{1}{2} m_s}(j) \eta_{\frac{1}{2} m_t}(j),
\end{equation} }
where $\chi_{\frac{1}{2} m_s}$ and \, $\eta_{\frac{1}{2} m_t}$ are,
respectively, the spin and isospin wave functions for $j$-th nucleon
~\cite{18}, and $\psi_{\mathbf{n}_i}$ are the one-particle orbitals. It
is clear that the success of such treatment will mainly depend on the
appropriate choice of variational basis for one-particle orbitals.
Conventionally, for the functions $\psi_{\mathbf{n}_i}$ , one employs the
functions of harmonic oscillator. In our model, which is also called in
Ref.~\cite{14} as "the model of polarized orbitals", each single orbital
is represented by an oscillator function as well. However, in contrast to
other authors, we use for every $i$-th one-particle state its own
independent variational parameters $(a_i,b_i,c_i)$.

Note that, as dictinct to the cluster models, all the nucleons equally
participate here in forming the dynamics of the system.

The optimal values for the variational parameters  $\{a_i,b_i,c_i\}$ as
well as the optimal configuration (determined by the set of quantum
numbers $\{\mathbf{n}_i=(n_i^x,n_i^y,n_i^z)\}$) are found as a result of
minimizing the functional $f_E(\{\mathbf{n}_i;a_i,b_i,c_i\})$ for the
total energy of the nucleus under consideration. If the chosen
variational basis of the functions $\psi_{\mathbf{n}_i}$ is flexible
enough, we can expect that the solution
$\Psi_{IM,\alpha}^{[f]}(1,2,...,A)$ to the many-particle problem found in
such way would be sufficiently realistic.

We are not going to dwell upon the so-called "technical" issues,
although, exactly these latter frequently determine the success of a
work. Let us only list the steps made for evaluating the functional
$f_E(\{\mathbf{n}_i;a_i,b_i,c_i\})$ :i) matrix elements for the operators
of all physical quantities entering the Hamiltonian (1) have been
obtained; ii) projection of the internal wave function (4) onto the
energy state $J^{\pi}=\frac{1}{2}^+$, associated with the ground state of
$^{11}$Be  has been carried out on the basis of the Hill-Wheeler
integral; iii) search for the global minimum of the functional
$f_E(\{\mathbf{n}_i;a_i,b_i,c_i\})$ for the total energy of the nucleus
in the hyperspace of variational parameters $\{a_i,b_i,c_i\}$ has been
performed.

\section{Results of numerical computations}

\qquad As mentioned above, the calculations of the needed wave function
$\Psi_{IM,\alpha}^{[f]}(1,2,...,A)$ and some basic features of the
nucleus $^{11}$Be have been performed in previous works ~\cite{14}.
Therefore, we can use here the results of ~\cite{14} in order to evaluate
the sought-for distribution of the nuclear matter in $^{11}$Be.

Recall that, according to [14], the ground state of the nucleus $^{11}$Be
is associated with the configuration:
 {\normalsize{
\begin{eqnarray} \label{5}
 \mbox{{\bf protons}  $ Z=4 \rightarrow(000)^2(001)^2 $;}\quad
 \mbox{{\bf neutrons} $ N=7 \rightarrow(000)^2(001)^2(010)^2(002)^1$},
\end{eqnarray}
 }}
i.e., the protons have the regular subsequently filled configuration,
whereas the neutron configuration is unusual. The optimal values for
variational parameters of the orbitals $\psi_{\mathbf{n}_i}$ are given in
Table 1.

Note that the physical meaning of the variational parameters
$(a_i,b_i,c_i)$ is that they determine the extent of spreading for the
i-th one-particle state along $x,y,z$ axis, respectively. The neutron
configuration indicated in (5) turned out to be energetically favorable
only in our model of "polarized"\, orbitals, where for the first time in
calculations of such kind the variational parameters $\{a_i,b_i,c_i\}$
were independent for any of the orbitals
 $\psi_{\mathbf{n}_i}$  and in any of the directions $x,y,z$.
Unproportionally large value of the variational parameter
 $c_4^n=5.38\, fm$ (see Table 1) evidences the anomalous spreading of the
valence neutron along one of the coordinate axis.

The information on the explicit expression for the wave function enables
us to evaluate the sought-for mass distribution
 $\varrho_A(\mathbf{r})$ in the nucleus  $^{11}$Be. To this end, we have
to average the operator for the one-particle spatial density with the
wave functions $\Psi_{IM,\alpha}^{[f]}(1,2,...,A)$ , i.e.
{\normalsize
\begin{equation}
 \label{6}
\varrho_A(\mathbf{r})=
\langle\Psi_{IM,\alpha}^{[f]}|\sum_{i=1}^{A}\delta(\mathbf{r}-\mathbf{r}_i)|\Psi_{IM,\alpha}^{[f]}\rangle
\,.
\end{equation} }
Here the coordinates  $\mathbf{r}_i$  are written in center mass system
$\mathbf{R}_{c.m.}=0$.

Note that in experiments, only the radial dependence for the nuclear
density distribution averaged over the angles  $\Omega$
{\normalsize
\begin{equation}
 \label{7}
\varrho_A(r)=\frac{1}{4\pi
r^2}\,\int_0^{2\pi}d\vartheta\int_0^{\pi}\varrho_A(\mathbf{r})\sin\theta
d\theta \, ; \quad  r>1 \mbox{\it fm}
\end{equation} }
is measured. In the vicinity of the point $r=0$, the averaging over the
angles $\Omega$ has to be performed as
{\normalsize
\begin{equation}
 \label{8}
\varrho_A(r)\approx\frac{1}{N_{points}}\,\sum_{i=1}^{N_{points}}\varrho_A(\mathbf{r}_i)
\, ; \quad r<2 \mbox{\it fm},
\end{equation} }
where the points  $\mathbf{r}_i$ are taken on the sphere of the radius
$r=|\mathbf{r}_i|$.

The distribution of nuclear matter for the wave function of the ground \,
$\frac{1}{2}^+$ state  of the nucleus calculated in the above way is
shown in Fig.1. Actually, this figure represents an almost identical copy
of Fig.2 of Ref.~\cite{8}, where, in addition, our theoretical curve (the
line with dots) is displayed. The three curves with $\beta=0$,
$\beta=0.5$ and $\beta=0.7$ in Fig.1 are the attempts of other authors to
describe the experimental situation within the framework of the deformed
shell model (see the references in ~\cite{8}). As is seen from Fig.1, the
theoretical data (the line with dots) and the experimental estimates for
the nuclear density distribution in the nucleus $^{11}$Be are in good
agreement. It is particularly true for the magnitude and the slope in the
"tail", where the experimental estimates are very reliable and a complete
coincidence between the theory and experiment is observed.

It is worth mentioning that in theory, in contrast to experimental
studies, there exists a possibility to investigate the above mass
distribution in the internal coordinate system rigidly coupled to the
nucleus. In this case we deal with the spatial structure
 $\varrho_A(\mathbf{r})$, which is the function of three coordinates
$x,y,z$. In Fig.2, we give one of the sections (in $(x,z)$ plane) for the
function $\varrho_A(\mathbf{r})$ for the nucleus  $^{11}$Be under
consideration. The figure in the plane $(y,z)$ would be similar, as far
as the size properties are concerned. The choice of the values for the
lines of constant density in Fig.2 has been made in agreement with the
relevant experimental data within the interval $[5 \mbox{\it fm} <r<12
\mbox{\it fm}]$. Finally, in Fig.3 we present the distribution of nuclear
matter $\varrho_{neutron}(\mathbf{r})$  for the valence nucleon, or, more
exactly, the more exactly, the lines of constant density in the plane
(x,z). In this figure, we clearly observe three structures, the two
lateral ones, symmetrically placed at the distance of $10 \mbox{\it fm}$,
and the central one being considerably smaller in size than the lateral
structures. All these three structures have approximately equal density
with the magnitude $(2-5)\cdot10^{-5} \mbox{\it fm$^{-3}$}$.

\section{Analysis and conclusions}

 \quad After introduction to both experimental and theoretical results on the
mass distribution in halo nucleus $^{11}$Be, let us try to make analysis
in order to obtain a physical interpretation. Let us start with Fig.1.
The radial distribution for the density of nuclear matter presented here
could be hardly brought into agreement with the notion of  "atomic
nucleus"\, conventionally given in reference books. Specifically, the
nucleon density is said therein to be nearly constant in the nucleus
center, and to decrease exponentially at periphery. As is seen from
Fig.1, the nucleus $^{11}$Be is not the case. Other features of nucleon
system under consideration are in contrast with the classical concept of
"atomic nucleus"\, as well, let alone the size properties.

Fig.2 and Fig.3 strongly evidence that in the case of the halo nucleus
$^{11}$Be we deal, in fact, with two distinctly different physical
systems, the well known nucleus $^{10}$Be referred here to as the "core
nucleus", and the structure of intermediate, between atomic and nuclear,
nature. The latter is presented by three separate objects. This structure
is, first of all, a result of the action of the Pauli principle with
respect to all 11 nucleons constituting the $^{11}$Be system. It is
coupled to the core nucleus by its central part, which completely
envelopes the core nucleus. Two lateral much more massive parts are
outside the nucleus $^{10}$Be, i.e., beyond the action of nuclear forces.
These three parts-objects are held together only due to the fact that
they originate from a single quantum of nuclear matter, the neutron.

To put it in the other way, the nuclear processes give rise to the
conditions, which make it energetically favorable for the system of 4
protons and 7 neutrons to form the compact nucleus $^{10}$Be and an
unusual structure with the mass and the quantum numbers being exactly
equal to the valence neutron. According to the experimental estimates,
this energy gain is equal to only $ \delta E=0.32\, MeV$. Exactly at this
distance, on energy scale, the first excited $\frac{1}{2}^-$ state of the
nucleon system under consideration is observed.

 Thus, the analysis of the experimental situation on the distribution of
nuclear matter in the exotic nucleus $^{11}$Be\, performed in this work
has shown that the nucleon system $^{11}$Be\, should be considered as two
different in nature subsystems, the compact nucleus $^{10}$Be, and some
structure emerging in the vicinity of this nucleus as a result of Pauli
principle, which is distinctly different in its features from an atomic
nucleus. Therefore, we should regard the discovery of the halo nuclei of
I-st kind as the discovery of new structures of microcosm with the sizes
ranging up to tens of $fm$ and the density being thousand times less than
the nuclear one.

 \mbox{}
\newpage
\begin{table}
\caption{ The optimal values of the variational  parameters for
one-particle orbitals  $\psi_{\mathbf{n}_i}$ in $^{11}$Be.}
 \centerline{}
 \begin{center}
 \begin{tabular}{|c|c|}        \hline
 Parameters $\{a_i, b_i, c_i\}.$           & The optimal values, $fm$.  \\   \hline \hline
                 &          \\
 $a_1^p$         &  1.355   \\
 $a_2^p$         &  1.49    \\   \hline
                 &          \\
 $b_1^p$         &  1.55    \\
 $b_2^p$         &  1.685   \\    \hline
                 &          \\
 $c_1^p$         &  1.84    \\
 $c_2^p$         &  1.92    \\    \hline \hline
                 &          \\
 $a_1^n$         &  1.375   \\
 $a_2^n$         &  1.51    \\
 $a_3^n$         &  1.61    \\
 $a_4^n$         &  1.725   \\     \hline
                 &          \\
 $b_1^n$         &  1.52    \\
 $b_2^n$         &  1.64    \\
 $b_3^n$         &  1.80    \\
 $b_4^n$         &  1.90    \\     \hline
                 &          \\
 $c_1^n$         &  1.89    \\
 $c_2^n$         &  1.96    \\
 $c_3^n$         &  2.18    \\
 $c_4^n$         &  5.38    \\     \hline
\end{tabular}
 \end{center}
\end{table}

\begin{figure}[p]
\centering
\includegraphics[width=4in]{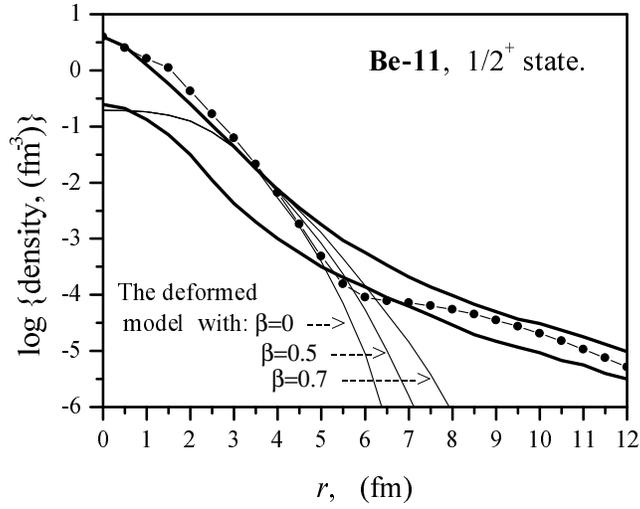}
\caption{Density distribution of nucleons on logarithmic scale for the
ground $\frac{1}{2}^+$  state of $^{11}$Be. The results of RIKEN
~\cite{8} are shown by thick lines; the results of our calculation are
shown by thin line with filled circles. The thin curves with  $\beta=0,\,
\beta=0.5$\,and $\beta=0.7$ are the results of the deformed model (see
~\cite{8}).
}
\label{fig1}
\end{figure}

\begin{figure}[p]
\centering
\includegraphics[width=4in]{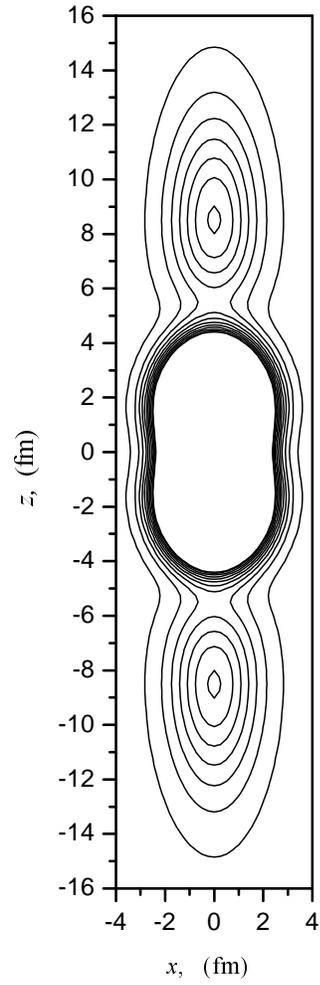}
\caption{Intrinsic structure of $\frac{1}{2}^+$  state of $^{11}$Be. The figure
shows  sections of the matter density $\varrho_{A}(x,y=0,z)$ in the plane
$(x,z)$.}
\label{fig2}
\end{figure}

\begin{figure}[p]
\centering
\includegraphics[width=4in]{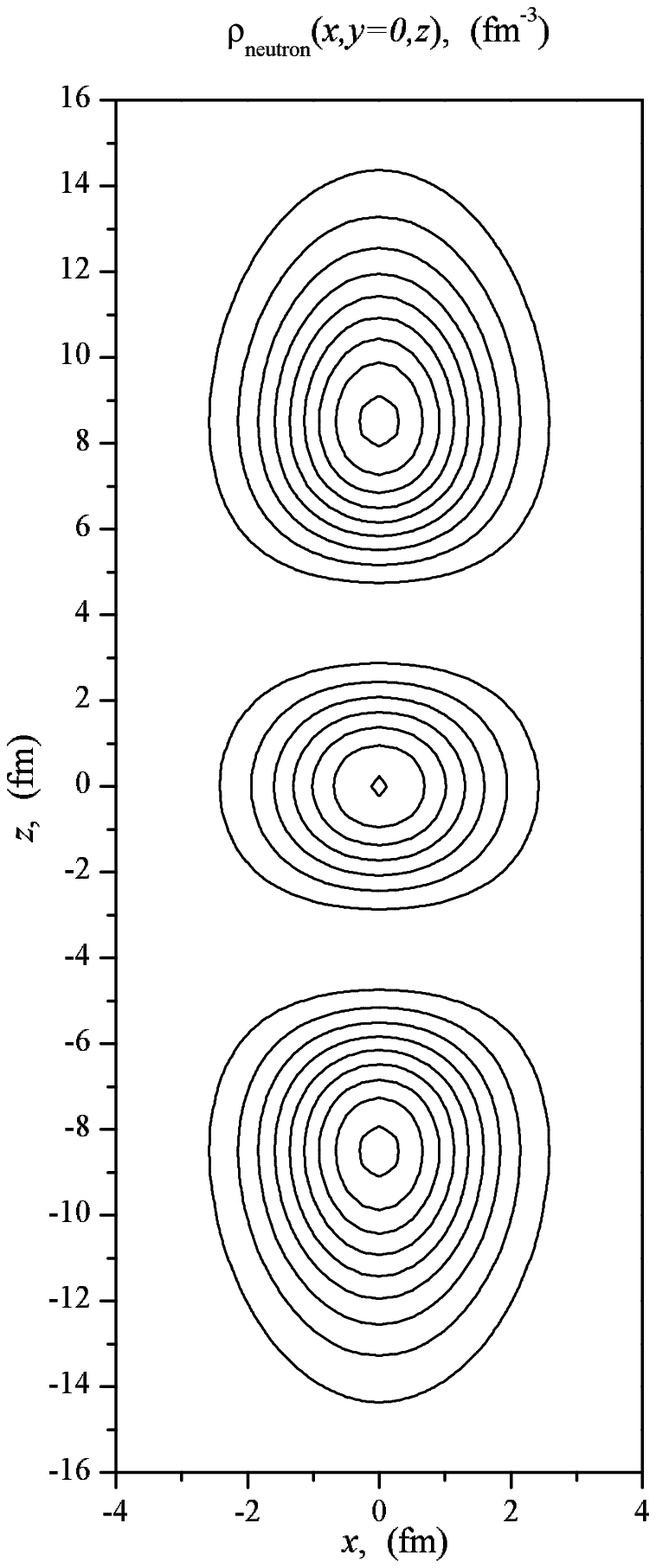}
\caption{The density distribution for the valence neutron in the plane
$(x,z)$.}
\label{fig3}
\end{figure}


\begin{thebibliography}{18}

\bibitem{1} I.Tanihata et al., Phys. Rev. Lett., {\bf55} (1985) 2676.

\bibitem{2}R.Kalpakchieva, Yu.E. Penionzhkevich, H.G. Bohlen, Fiz. Elem. Chastits At. Yadra, {\bf 29} (1998) 832.

\bibitem{3} R.Kalpakchieva, Yu.E.Penionzhkevich, H.G. Bohlen, Fiz. Elem. Chastits At. Yadra, {\bf 30} (1999) 1429.

\bibitem{4} I.Tanihata, Prog. Part. Nucl. Phys., {\bf35} (1995) 505.

\bibitem{5}I.Tanihata, J. Phys. G., Nucl. Phys., {\bf22} (1996) 157.

\bibitem{6} K.Riisager, Rev. Mod. Phys., {\bf66} (1994) 1105.

\bibitem{7} M.V.Zhukov et al., Phys. Rep.,  {\bf231} (1993) 151.

\bibitem{8} M.Fukuda et al.,  Phys.Lett., {\bf B268} (1991) 339.

\bibitem{9} J.S.Al-Khalilil, J.A.Tostevin, Phys.Rev.Lett.,  {\bf76}, p.3903 (1996).

\bibitem{10}J.S.Al-Khalilil et al., Phys.Rev.,  {C\bf54} (1996) 1843.

\bibitem{11} T.Nakamura et al., Phys.Lett., {B\bf331} (1994) 296.

\bibitem{12}  H.Esbensen et al., Phys.Rev., v.{C\bf56} (1997) 3054.

\bibitem{13} K.Wildermuth, Y.C.Tang. A unified theory of the nucleus, Vieweg,  Braunschweig, 1977.

\bibitem{14} A.I.Steshenko, Yad. Fiz.,{\bf60} (1997) 599.

\bibitem{15} I.V.Simenog and A.I.Steshenko, Ukrainian Phys.Journal (in ukrainian), {\bf38} (1993) 381.

\bibitem{16} S. De Benedetti, Nuclear Interactions, John Wiley and Sons, Inc., New York-London-Sydney, 1964.

\bibitem{17} A.S.Davydov, Quantum Mechanics, Ì.,"Nauka", 1973, \S51.

\bibitem{18} O.F.Nemec, V.G.Neudatchin, A.T.Rudchik, Yu.F.Smirnov, Yu.M.Tchu-vil'sky,
       Nucleonic clusters in atomic nuclei and manybody transfer nuclear reactions.
       Naukova Dumka, Kiev, 1988.

\end{thebibliography}
\end{document}